\documentclass{aa}
\usepackage{graphics,epsfig,times}

\usepackage{natbib}
\bibpunct{(}{)}{;}{a}{}{,}

\newcommand{\oergs}[1]{$10^{#1}$ erg s$^{-1}$}
\newcommand{\hcm}[1]{$\times 10^{#1}$ cm$^{-2}$}
\newcommand{\ohcm}[1]{$10^{#1}$ cm$^{-2}$}
\newcommand{\expo}[1]{$\times 10^{#1}$}
\newcommand{\oexpo}[1]{$10^{#1}$}
\newcommand{\nh}{N$_{\rm H}$}

\newcommand{\lx}{\hbox{L$_{\rm x}$}}
\newcommand{\fx}{\hbox{F$_{\rm x}$}}
\newcommand{\lasc}{\hbox{L$_{\rm x,ASCA}$}}
\newcommand{\lros}{\hbox{L$_{\rm x,ROSAT}$}}
\newcommand{\lmax}{\hbox{L$_{\rm x}^{\rm max}$}}
\newcommand{\lmin}{\hbox{L$_{\rm x}^{\rm min}$}}

\newcommand{\Halp}{H${\alpha}$}
\newcommand{\ltsima}{$\buildrel < \over \sim$}
\newcommand{\lsim}{\lower.5ex\hbox{\ltsima}}
\newcommand{\gtsima}{$\buildrel > \over \sim$}
\newcommand{\gsim}{\lower.5ex\hbox{\gtsima}}

\newcommand{\sax}{\hbox{\object{SAX\,J0103.2$-$7209}}}
\newcommand{\snra}{\hbox{\object{SNR\,0047$-$73.5}}}
\newcommand{\snrb}{\hbox{\object{SNR\,0103$-$72.6}}}
\begin{document}
 
\title{X-ray observations of Be/X-ray binaries in the SMC
        \thanks{Based on observations with XMM-Newton,
               an ESA Science Mission with instruments and contributions 
               directly funded by ESA Member states and the USA (NASA)}}
 
%\author{F.~Haberl\inst{1} \and W.~Pietsch\inst{1} \and ?}
\author{F.~Haberl \and W.~Pietsch}

\titlerunning{X-ray observations of Be/X-ray binaries in the SMC}
\authorrunning{Haberl \& Pietsch}
 
\offprints{F. Haberl, \email{fwh@mpe.mpg.de}}
 
\institute{Max-Planck-Institut f\"ur extraterrestrische Physik,
           Giessenbachstra{\ss}e, 85748 Garching, Germany}
 
\date{Received 14 July 2003 / Accepted 21 October 2003}
 
\abstract{
Fifteen Be/X-ray binaries and candidates in the SMC were observed serendipitously with the EPIC 
instruments of XMM-Newton during two observations of SNR\,0047-73.5 and SNR\,0103-72.6 in 
October 2000. A total of twelve of those sources are detected. For eleven of them an accurate 
position and in part detection of X-ray pulsations support the proposed identification 
as Be/X-ray binaries.
In one case the improved X-ray position excludes the previously suggested identification with an
\Halp\ emission line star found within the ROSAT error circle. The detection of pulsations 
(172.2 s, 320.1 s and 751 s) from three hard X-ray sources with periods known from ASCA 
observations confirm their proposed identifications with ROSAT sources and their optical Be 
star counterparts. In addition, pulsations with a period of 263.6 s were found from 
XMMU\,J004723.7$-$731226=RX\,J0047.3$-$7312. For \sax\ a pulse period of 341.2$\pm$0.5 s 
was determined, continuing the large spin-up seen with ASCA, BeppoSAX and Chandra between 
1996 and 1999 with a period derivative of $-$1.6 s yr$^{-1}$ covering now 4.5 years. The 
0.3-10.0 keV EPIC spectra of 
all eleven Be/X-ray binaries and candidates are consistent with power-law energy distributions 
with derived photon indices strongly peaked at 1.00 with a standard deviation of 0.16. No 
pulsations are detected from RX\,J0049.2$-$7311 and RX\,J0049.5$-$7310 (both near the 9 s pulsar 
AX\,J0049$-$732) and  RX\,J0105.1$-$7211 (near AX\,J0105$-$722, which may pulsate with 3.3 s), 
leaving the identification of the ASCA sources with ROSAT and corresponding XMM-Newton 
objects still unclear. We present an updated list of high mass X-ray binaries (HMXBs) and 
candidates in the SMC incorporating improved X-ray positions obtained from Chandra and
XMM-Newton observations. Including the results from this work and recent publications the
SMC HMXB catalogue comprises 65 objects with at least 37 showing X-ray pulsations.

}
 
\maketitle
 
\section{Introduction}

X-ray observations of the Small Magellanic Cloud (SMC) with recent observatories
like ASCA, BeppoSAX, ROSAT and RXTE revealed a large number of Be/X-ray binary systems.
Most of them were discovered through the detection of X-ray pulsations
\citep[e.g. see summary of ASCA X-ray sources in the SMC][]{2003PASJ...55..161Y}, indicating
the spin period of the neutron star which orbits an early type star. Optical follow-up
observations are then required to determine the spectral type of the high mass star
and in the SMC only Be stars were identified (with the only secure exception of the supergiant
system SMC\,X-1), making the SMC very different to the Milky Way and the 
Large Magellanic Cloud in absolute numbers of high mass X-ray binaries with respect to
the galaxy mass and also in their ratio of Be to supergiant OB systems. 
 
\citet[][ hereafter HS00]{2000A&A...359..573H} searched for Be/X-ray binaries in the ROSAT
catalogues of \citet{2000A&AS..142...41H} and \citet{2000A&AS..147...75S} by correlating
them with catalogues of optical \Halp\ emission line objects 
\citep{1993A&AS..102..451M,2000MNRAS.311..741M}. HS00 presented a list of 47 Be/X-ray
binaries and candidates with X-ray positions typically better than 10\arcsec. Since 
the launch of Chandra and XMM-Newton new X-ray pulsars and Be/X-ray binary candidates
were discovered \citep{2001A&A...369L..29S,2003A&A...403..901S,2003ApJ...584L..79M}. 
Together with eleven pulsars detected by RXTE with very uncertain X-ray positions we 
know about sixty high mass X-ray binaries (HMXBs) and candidates in the SMC, from 
which more than thirty are known as X-ray pulsars 
\citep{2003PASJ...55..161Y,2001foap.conf..239M}.

The position uncertainties of X-ray sources in the SMC obtained by previous 
missions of up to tens of arc-seconds (including the most precise ROSAT positions) made it in many 
cases impossible to uniquely correlate sources from the different X-ray missions, 
nor allowed exact optical identifications. We therefore 
started to systematically analyse archival XMM-Newton data of HMXBs and candidates 
in the SMC in order to confirm their proposed nature. First results were presented
in \citet{2003A&A...403..901S}. Here we continue this work analyzing the XMM-Newton 
EPIC data of two fields in the SMC which contain in total fifteen known HMXBs and 
candidates.

\section{XMM-Newton observations and analysis}

XMM-Newton observed the fields around the SMC supernova remnants \snra\ and 
\snrb\ on October 15, 2000 and on October 17, 2000 with EPIC-MOS 
\citep[MOS1 and MOS2,][]{2001A&A...365L..27T} for 26.9 ks and 34.8 ks and EPIC-pn 
\citep[][]{2001A&A...365L..18S} for 23.0 ks and 30.9 ks. For both observations 
the three EPIC cameras were operated in imaging mode covering
the full field of view of $\sim$14\arcmin\ radius, MOS in full-frame 
(2.6 s time resolution) and pn in extended full-frame (0.2 s) CCD readout mode.
For optical light blocking the medium filter was used in all cameras.

The data were processed using the XMM-Newton analysis package SAS version 5.4.1 to 
produce the photon event files and binned data products like images, spectra and 
light curves. To avoid artificial photon arrival time jumps in the EPIC pn data an 
updated SAS development version from 2003, May 16 was used to create pn event files.

\begin{figure}
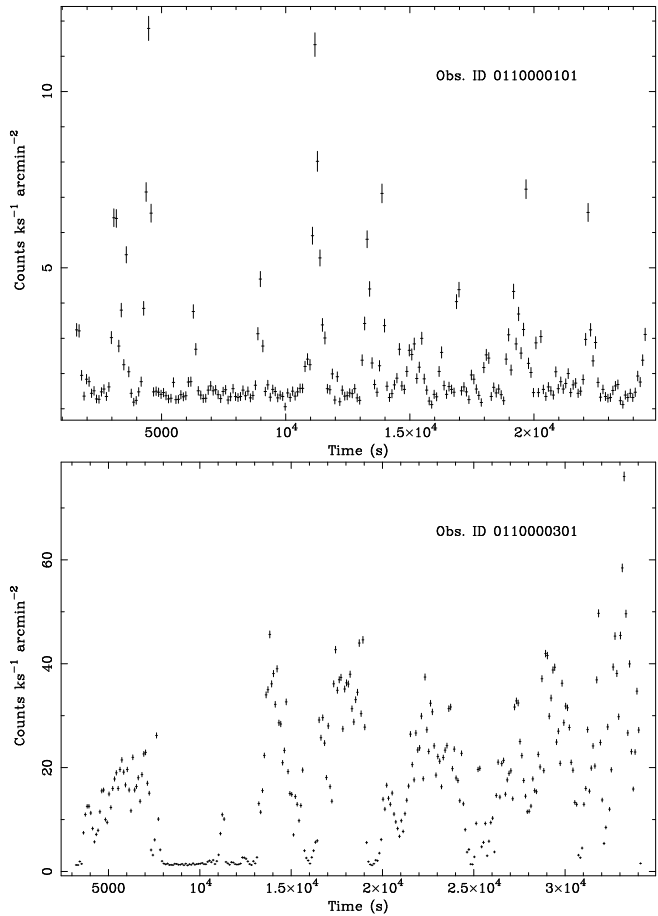

\begin{center}
\resizebox{8.6cm}{!}{\includegraphics[clip=,angle=-90]{H4667_p0110000101pns005fbktsr0000_lcurve.ps}}
\resizebox{8.6cm}{!}{\includegraphics[clip=,angle=-90]{H4667_p0110000301pns005fbktsr0000_lcurve.ps}}
\end{center}
\caption{EPIC-pn background light curves obtained from source free regions in 
the 7.0-15.0 keV band for the two investigated observations of \snra\ (0110000101) 
and \snrb\ (0110000301).}
\label{fig-backgr}
\end{figure}

\begin{figure}
\begin{center}
\resizebox{8.3cm}{!}{\includegraphics[clip=]{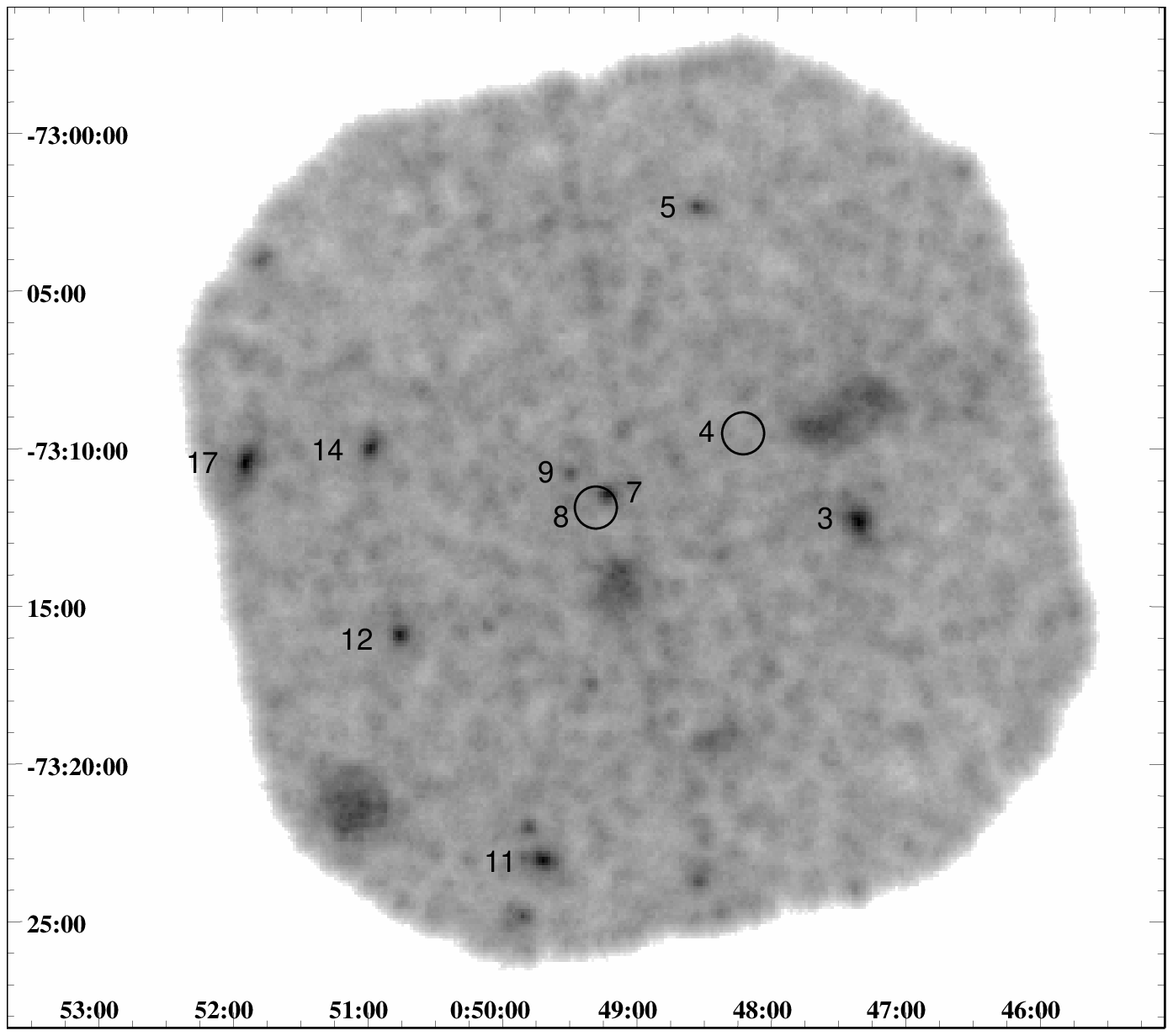}}
\resizebox{8.3cm}{!}{\includegraphics[clip=]{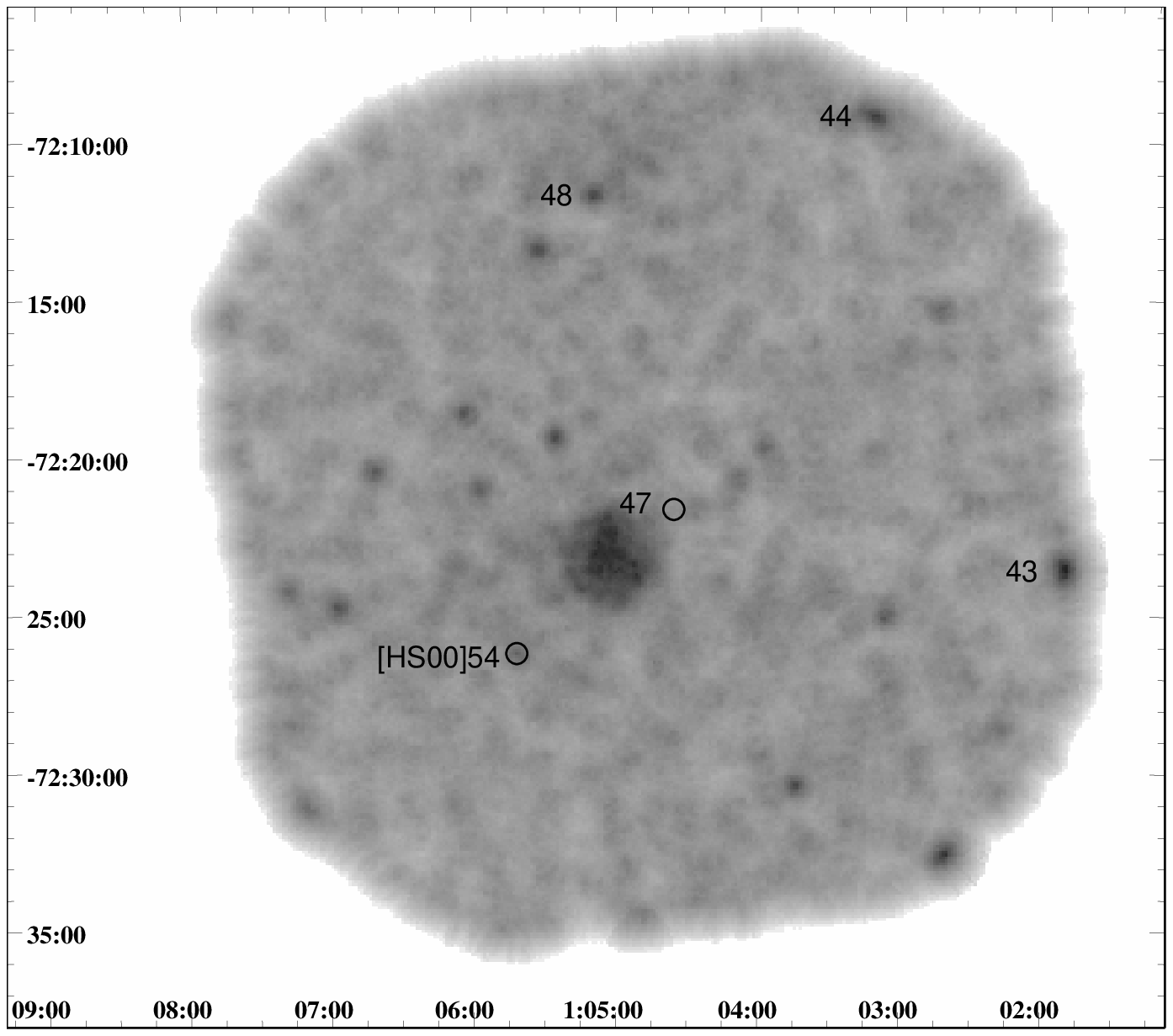}}
\end{center}
\caption{EPIC-pn images of the two observed SMC fields obtained from 1.0-2.0 keV data, using an 
intensity dependent adaptive smoothing (top: observation 0110000101 with \snra\ near the center 
of the image; bottom: observation 0110000301 with \snrb). 
The Be/X-ray binaries and candidates are marked with 
numbers as used in Table~\ref{tab-smc}. Source 54 from HS00 was removed as Be/X-ray binary 
candidate from Table~\ref{tab-smc} because its improved position derived from the EPIC data is now 
inconsistent with that of the emission line object proposed as optical counterpart. 
Sources 4, 8 and 47 were not detected,
while a possible association of source 8 (detected with ASCA) with source 7 or 9 (ROSAT) could not 
be clarified (see Sect.~2.1).}
\label{fig-ima}
\end{figure}

The observation of \snrb\ suffered from strong background flaring activity 
(Fig.~\ref{fig-backgr}) which hampered the analysis of the data. We applied
two different levels of background screening depending on the goals of the 
analysis. To obtain clean images which also show faint sources we used only data
from times of low background ($<$10 cts ks$^{-1}$ arcmin$^{-2}$ in EPIC-pn). 
To search for periodicities in the brighter sources we used two different background cuts
of 10 and 50 cts ks$^{-1}$ arcmin$^{-2}$. While the strong screening removes the flares
which have typical time scales of a few 1000 s, the latter method avoids too many data gaps. 
For the spectral analysis of point sources the cut level of 50 cts ks$^{-1}$ arcmin$^{-2}$
was applied. For the MOS data corresponding levels of 2 and 5 cts ks$^{-1}$ arcmin$^{-2}$
were defined, which result in similar good time intervals in the contemporaneous parts of 
MOS and pn data.

\begin{table*}
\label{tab-detect}
\begin{center}
\caption[]{Be/X-ray binaries and candidates detected in the EPIC fields.}
\begin{tabular}{lllcrrrrcc}
\hline\noalign{\smallskip}
\hline\noalign{\smallskip}
\multicolumn{1}{l}{Column 1} &
\multicolumn{1}{c}{2} &
\multicolumn{1}{c}{3} &
\multicolumn{1}{c}{4} &
\multicolumn{1}{c}{5} &
\multicolumn{1}{c}{6} &
\multicolumn{1}{c}{7} &
\multicolumn{1}{c}{8} &
\multicolumn{1}{c}{9} &
\multicolumn{1}{c}{10} \\

\noalign{\smallskip}\hline\noalign{\smallskip}
\multicolumn{1}{l}{XMMU\,J...} &
\multicolumn{1}{c}{RA} &
\multicolumn{1}{c}{Dec} &
\multicolumn{1}{c}{Pos. err.} &
\multicolumn{1}{c}{Count rate} &
\multicolumn{1}{c}{HS} &
\multicolumn{1}{c}{MA} &
\multicolumn{1}{c}{d$_{\rm MA}$} &
\multicolumn{1}{c}{Period} &
\multicolumn{1}{c}{Source} \\

\multicolumn{1}{c}{} &
\multicolumn{2}{c}{(J2000.0)} &
\multicolumn{1}{c}{[\arcsec]} &
\multicolumn{1}{c}{[s$^{-1}$]} &
\multicolumn{1}{c}{} &
\multicolumn{1}{c}{} &
\multicolumn{1}{c}{[\arcsec]} &
\multicolumn{1}{c}{[s]} &
\multicolumn{1}{c}{Tab.\,\ref{tab-smc}} \\

\noalign{\smallskip}\hline\noalign{\smallskip}
004723.7$-$731226 & 00 47 23.70 & $-$73 12 26.9 & 4.0 & 4.14\expo{-1}$\pm$6\expo{-3} &  6 &  172   &  3.9 & 263.64$\pm$0.30 & ~3 \\
004834.5$-$730230$^a$ & 00 48 34.50 & $-$73 02 30.0 & 4.1 & 5.49\expo{-2}$\pm$5\expo{-3} &  8 &  238   &  2.2 &	            & ~5 \\
004913.8$-$731136 & 00 49 13.84 & $-$73 11 36.7 & 4.0 & 3.74\expo{-2}$\pm$2\expo{-3} & -- &   --   &   -- &	            & ~7 \\
004929.9$-$731058 & 00 49 29.92 & $-$73 10 58.0 & 4.1 & 8.37\expo{-3}$\pm$8\expo{-4} & 10 &  300   &  1.7 &	            & ~9 \\
004942.3$-$732313 & 00 49 42.37 & $-$73 23 13.2 & 4.0 & 1.85\expo{-1}$\pm$5\expo{-3} & 12 &  315   &  2.6 & 750.9$\pm$2.4   & 11 \\
005045.2$-$731602 & 00 50 45.20 & $-$73 16 02.9 & 4.0 & 9.52\expo{-2}$\pm$3\expo{-3} & 13 &  387   &  3.0 & 320.12$\pm$0.40 & 12 \\
005057.6$-$731007 & 00 50 57.60 & $-$73 10 07.9 & 4.0 & 8.40\expo{-2}$\pm$3\expo{-3} & 15 &  414   &  1.7 &	            & 14 \\
005152.2$-$731033 & 00 51 52.29 & $-$73 10 33.4 & 4.0 & 3.41\expo{-1}$\pm$7\expo{-3} & 19 &  504   &  2.1 & 172.21$\pm$0.13 & 17 \\
\noalign{\smallskip}\hline\noalign{\smallskip}
010152.4$-$722335 & 01 01 52.41 & $-$72 23 35.7 & 4.1 & 1.19\expo{-1}$\pm$7\expo{-3} & 45 & 1288   &  2.8 &	            & 43 \\
010314.2$-$720914 & 01 03 14.21 & $-$72 09 14.5 & 4.1 & 8.44\expo{-2}$\pm$7\expo{-3} & 49 & 1367   &  1.1 & 341.21$\pm$0.50 & 44 \\
010509.7$-$721146 & 01 05 09.76 & $-$72 11 46.4 & 4.2 & 2.06\expo{-2}$\pm$3\expo{-3} & 53 & (1517) & 10.7 &	            & 48 \\
010541.5$-$722617 & 01 05 41.56 & $-$72 26 17.3 & 4.2 & 3.52\expo{-3}$\pm$5\expo{-4} & 54 & (1544) & 14.3 &	            & -- \\
\noalign{\smallskip}\hline\noalign{\smallskip}
\end{tabular}
\end{center}
$^a$ Position and count rate from combined MOS data\\
Notes to specific columns: (5) 0.3-7.5 keV count rates with 1$\sigma$ errors. (6) Entry numbers 
from the compilation of Be/X-ray binaries from \citet{2000A&A...359..573H}. (7) Entry numbers from the \Halp\ emission 
line object catalogue of \citet{1993A&AS..102..451M}. (8) Angular distance between X-ray and the optical position of the nearest
emission line object. (9) Pulse period determined in this work.
\end{table*}

Fig.~\ref{fig-ima} shows the EPIC-pn images of the two observed SMC fields. The 
investigated Be/X-ray binaries are marked with source numbers given in Table~\ref{tab-smc}
which we use throughout the paper. The fields of view of the MOS cameras are slightly 
shifted with respect to pn and do not cover the 345 s pulsar \sax\ (source 44).

\subsection{Source positions}

Source detection based on sliding window and maximum likelihood methods available 
in the SAS package was applied to the MOS and pn images, simultaneously for four 
energy bands (0.3-1.0 keV, 1.0-2.0 keV, 2.0-4.5 keV and 4.5-7.5 keV), but separately 
for the three instruments. Dividing the data into several energy bands
enhances the signal to noise ratio for sources which have their major part of emission
in a relatively narrow band, like e.g. super-soft X-ray sources or highly absorbed
sources.
In Table~1 %Table~\ref{tab-detect}
we list the detected Be/X-ray binaries with XMM-Newton names, X-ray 
positions, count rates as obtained from the pn images 
(except for source 5 which is located near a CCD border in the pn camera). Coordinate 
uncertainties include a 4\arcsec\ systematic error \citep{2002A&A...382..522B}. 
We correlated our source list with the compilation of Be/X-ray binaries from HS00 
and with the \Halp\ emission line object catalogue of \citet{1993A&AS..102..451M}. 
Source 47 (Fig.~\ref{fig-ima}) 
detected in ROSAT HRI data \citep{2000A&AS..147...75S} was not 
detected by the EPIC instruments. Also source 4 
\citep{2003PASJ...55..161Y} was not detected. The association of 
the 9.1 s pulsar AX\,J0049$-$732 (source 8) with either of source 7 or 9 
\citep{2003PASJ...55..161Y,2000A&A...361..823F} could not be 
confirmed by a detection of the pulsations from either of the two sources 
in the EPIC data (see below). AX\,J0049$-$732 may be an uncorrelated
source detected during a brighter state during the ASCA observations.
HS00 classified RX\,J0105.7$-$7226 ([HS00]54) as candidate Be/X-ray binary due to the 
positional coincidence with the emission line object 1544 in \citet{1993A&AS..102..451M}. 
Our newly derived X-ray position fully agrees with the ROSAT position, but the smaller 
uncertainty now excludes an association with the emission line object. Since no other evidence 
argues for RX\,0105.7-7226 as Be/X-ray binary, we remove it from the list
of candidates. Similarly, source 48 is now very unlikely to be associated with [MA93]1517. 
This makes the situation concerning the identification of the possible pulsar 
AX\,J0105$-$722 with a ROSAT source more unclear. However,
the spectral analysis of RX\,J0105.1$-$7211 (see below) supports the HMXB nature of 
the source and we therefore keep it as a candidate.

There remain eleven Be/X-ray binaries and candidates in the two SMC fields for which 
we present a more detailed temporal and spectral analysis in the following.

\begin{table*}
\begin{center}
\caption[]{Spectral fit results using an absorbed power-law model.}
\begin{tabular}{llccrrrr}
\hline\noalign{\smallskip}
\hline\noalign{\smallskip}
\multicolumn{1}{l}{Source} &
\multicolumn{1}{c}{$\gamma$} &
\multicolumn{1}{c}{\nh$^{c}$} &
\multicolumn{1}{c}{\fx$^{d}$} &
\multicolumn{1}{c}{\lx$^{e}$} &
\multicolumn{1}{c}{\lasc$^{e}$} &
\multicolumn{1}{c}{\lros$^{e}$} &
\multicolumn{1}{c}{\lmax/\lmin} \\

\multicolumn{1}{l}{Tab.\,\ref{tab-smc}} &
\multicolumn{1}{c}{} &
\multicolumn{1}{c}{[\oexpo{21}cm$^{-2}$]} &
\multicolumn{1}{c}{[erg cm$^{-2}$ s$^{-1}$]} &
\multicolumn{1}{c}{[erg s$^{-1}$]} &
\multicolumn{1}{c}{[erg s$^{-1}$]} &
\multicolumn{1}{c}{[erg s$^{-1}$]} &
\multicolumn{1}{l}{} \\

\noalign{\smallskip}\hline\noalign{\smallskip}
~3       & 0.76$\pm$0.04          & 2.2$\pm$0.3	        & 3.3\expo{-12} & 1.8\expo{36} & 3.4-6.8\expo{35} & 1.5-12\expo{35}  & 12.0 \\
~5       & 0.89$\pm$0.19          & $<$1.7	        & 2.1\expo{-13} & 1.0\expo{35} &                  & 3.0\expo{35}     &  3.0 \\
~7       & 1.33$\pm$0.15          & 7.7$\pm$1.7	        & 2.7\expo{-13} & 1.8\expo{35} &                  & 1.6-2.0\expo{35} &  1.3 \\
~9$^{a}$ & 1.05$^{+0.55}_{-0.43}$ & 5.4$^{+6.3}_{-3.5}$ & 9.4\expo{-14} & 5.6\expo{34} &                  & 2.7-2.8\expo{35} &  4.8 \\
11       & 0.89$\pm$0.08          & 3.6$\pm$0.6	        & 1.6\expo{-12} & 9.0\expo{35} & 6.8-7.7\expo{35} & 1.0\expo{35}     &  9.0 \\
12       & 0.89$\pm$0.09          & 2.8$\pm$0.7	        & 5.5\expo{-13} & 3.1\expo{35} & 1.4-3.2\expo{36} & 2.5-3.6\expo{35} & 12.8 \\
14       & 0.90$\pm$0.09          & 2.4$\pm$0.7	        & 6.6\expo{-13} & 3.6\expo{35} & 4.5\expo{35}     & 2.4-3.6\expo{35} &  1.9 \\
17$^{b}$ & 1.03$\pm$0.08          & 0.6$\pm$0.3	        & 2.1\expo{-12} & 1.1\expo{36} & 3.0-6.8\expo{35} & 5.7-6.9\expo{35} &  3.7 \\
\noalign{\smallskip}\hline\noalign{\smallskip}
43       & 1.23$\pm$0.15          & 3.2$\pm$0.9	        & 9.7\expo{-13} & 5.8\expo{35} & 2.3-2.9\expo{35} & 4.0-5.8\expo{35} &  2.5 \\
44$^{b}$ & 1.03$\pm$0.28          & $<$1.7	        & 7.2\expo{-13} & 3.8\expo{35} & 3.6-15\expo{35\,f}  & 6.5-8.6\expo{35} &  4.2 \\
48       & 0.97$\pm$0.48          & $<$4.5	        & 1.4\expo{-13} & 7.7\expo{34} &                  & 1.6-3.1\expo{35} &  4.0 \\
\noalign{\smallskip}\hline\noalign{\smallskip}
\end{tabular}
\end{center}
$^a$ spectral fit to MOS spectra only; $^b$ spectral fit to pn spectrum only;\\
$^c$ total column density including galactic foreground column, assuming solar element abundances;\\
$^d$ 0.2--10 keV, determined from the pn spectrum;\\
$^e$ 0.2--10 keV, intrinsic luminosity with \nh\ set to 0, assuming a distance of 65 kpc to the SMC;\\
$^f$ the maximum luminosity was seen during a BeppoSAX observation;
\label{tab-fits}
\end{table*}

\subsection{X-ray pulsations}

\begin{figure*}
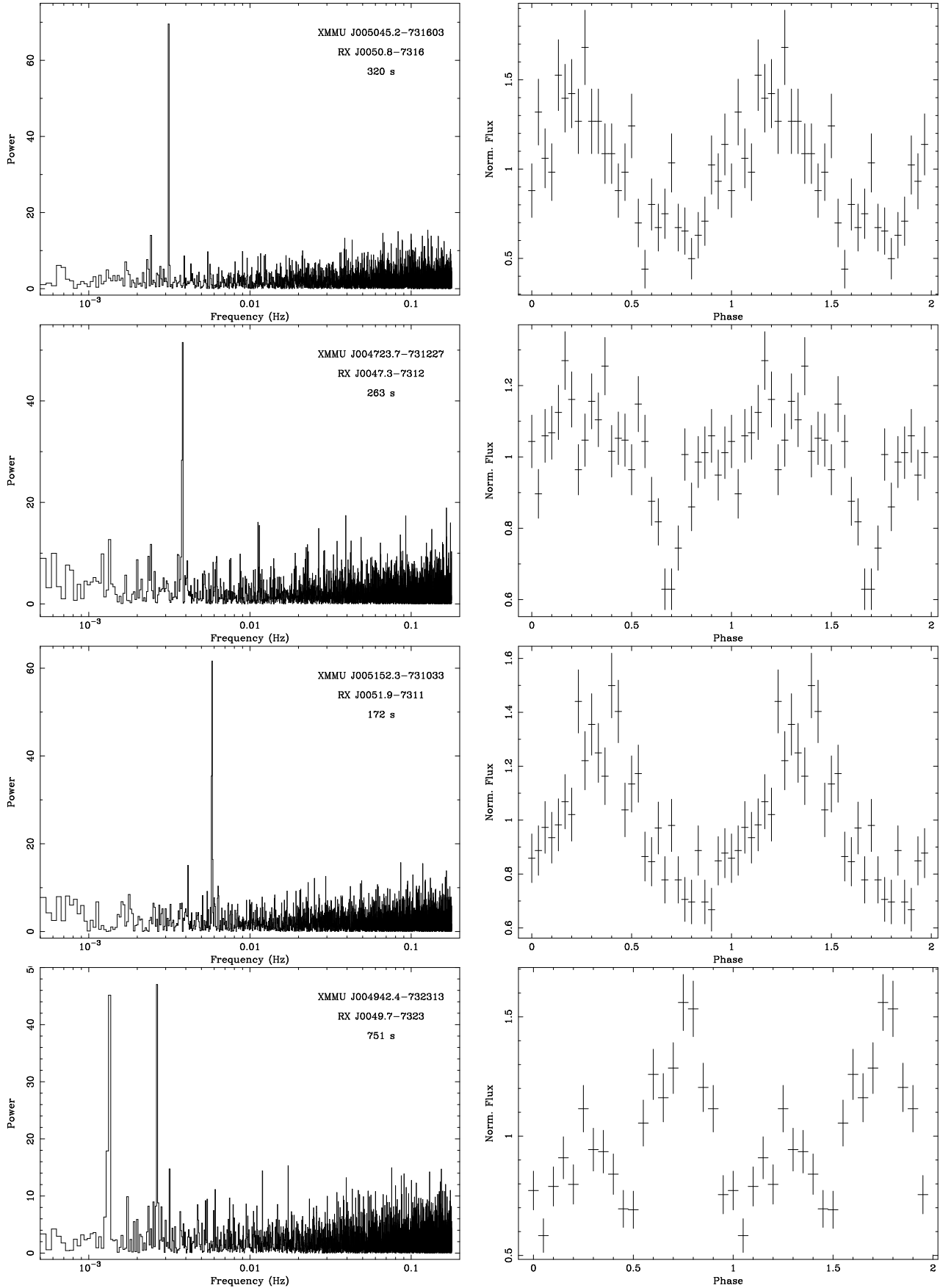

\begin{center}
\hbox{
\resizebox{8.2cm}{!}{\includegraphics[clip=,angle=-90]{H4667_pspc444_powspec.ps}}
\hspace{3mm}
\resizebox{8.2cm}{!}{\includegraphics[clip=,angle=-90]{H4667_pspc444_efold.ps}}
}
\hbox{
\resizebox{8.2cm}{!}{\includegraphics[clip=,angle=-90]{H4667_pspc434_powspec.ps}}
\hspace{3mm}
\resizebox{8.2cm}{!}{\includegraphics[clip=,angle=-90]{H4667_pspc434_efold.ps}}
}
\hbox{
\resizebox{8.2cm}{!}{\includegraphics[clip=,angle=-90]{H4667_pspc424_powspec.ps}}
\hspace{3mm}
\resizebox{8.2cm}{!}{\includegraphics[clip=,angle=-90]{H4667_pspc424_efold.ps}}
}
\hbox{
\resizebox{8.2cm}{!}{\includegraphics[clip=,angle=-90]{H4667_pspc468_powspec.ps}}
\hspace{3mm}
\resizebox{8.2cm}{!}{\includegraphics[clip=,angle=-90]{H4667_pspc468_efold.ps}}
}
\end{center}
\caption{Power spectra (left) and pulse-folded light curves (right) of X-ray 
pulsars in the EPIC field of \snra\ obtained from 0.3$-$7.5 keV pn data.}
\label{fig-pow}
\end{figure*}

We searched for pulsations in the X-ray light curves (from different energy bands and 
background screening levels) of all eleven Be/X-ray binaries and candidates using Fast 
Fourier Transform (FFT) and Rayleigh Z$_n^2$ methods. The power spectra obtained from 
broad band data (0.3$-$7.5 keV) of the three known pulsars in the field of \snra\ 
(sources 11, %: RX\,J0049.7$-$7323 = AX\,J0049.5$-$7323
 12 %: RX\,J0050.8$-$7316 = AX\,J0051$-$733,
 and 
 17) %: RX\,J0051.9$-$7311 = AX\,J0051.6$-$7311)
clearly reveal their pulse periods. This - together with the accurate source positions
- now allows the unique identification of the ASCA sources (which showed pulsations) 
with ROSAT sources (good X-ray positions), which could previously be done on 
positional coincidence only. In particular RX\,J0051.9$-$7311 (source 17), which 
was suggested to be the counterpart of the 16.6 s pulsar XTE\,J0050$-$732\#1 
\citep{2002ApJ...567L.129L} or the 172 s pulsar AX\,J0051.6$-$7311 
\citep{2003PASJ...55..161Y}, is consistent in position with 
XMMU\,J005152.2$-$731033, which shows 172.2 s pulsations. 
In addition we find 263 s pulsations from XMMU\,J004723.7$-$731226=RX\,J0047.3$-$7312
which is likely associated with AX\,J0047.3$-$7312 
\citep[see also][]{2003PASJ...55..161Y}. In Fig.~\ref{fig-pow} we present FFT 
power spectra and light curves folded at the pulse period of the four pulsars. 
As indicated by the presence of the first harmonic in the power spectrum of 
RX\,J0049.7$-$7323, the pulse profile of this long period pulsar is double 
peaked. The determined pulse periods with 1$\sigma$ error are listed in Table~1. %Table~\ref{tab-detect}

\begin{figure}
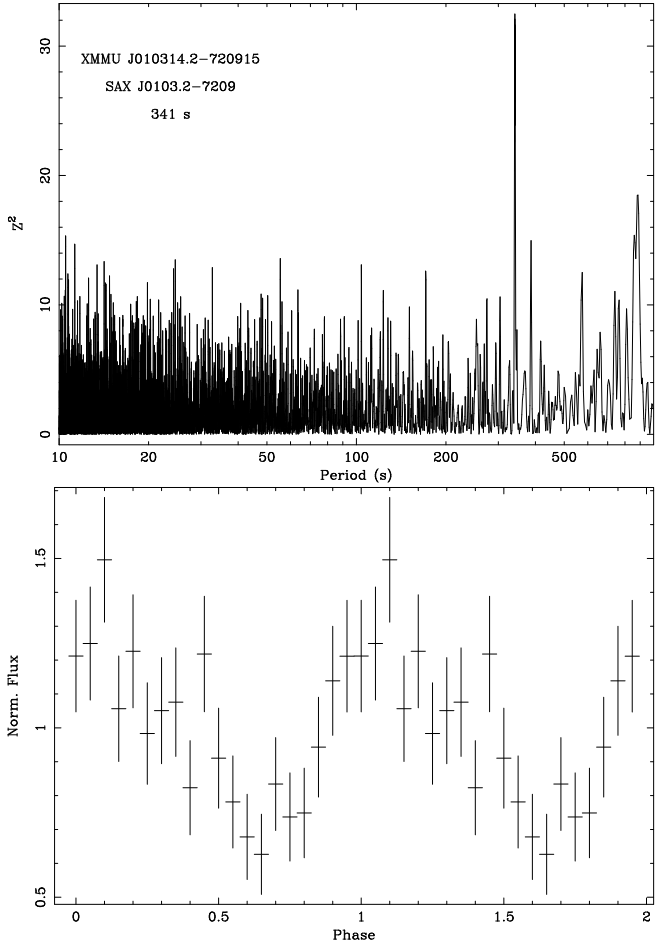

\begin{center}
\resizebox{8.6cm}{!}{\includegraphics[clip=,angle=-90]{H4667_pspc143_z2.ps}}
\resizebox{8.6cm}{!}{\includegraphics[clip=,angle=-90]{H4667_pspc143_efold.ps}}
\end{center}
\caption{Top: Periodogram resulting from a Z$^2_1$ test using the EPIC-pn data 
in the energy band 0.5$-$5.0 keV. Bottom: Folded pulse profile of the same data.}
\label{fig-spin}
\end{figure}

\begin{figure}
\begin{center}
\resizebox{8.6cm}{!}{\includegraphics[clip=,angle=-90]{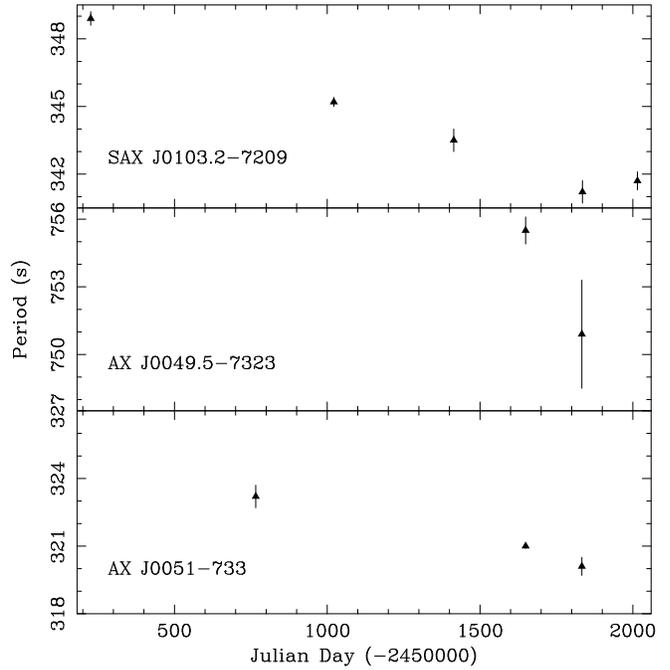}}
\end{center}
\caption{Spin history of three long period SMC pulsars. The last data points (for \sax\ 
the last two points with the very last taken from \citep{2003A&A...403..901S}) 
are XMM-Newton measurements. Previous values for the period were reported in  
\citet{2000ApJ...531L.131I} and \citet{2003PASJ...55..161Y}.}
\label{fig-hist}
\end{figure}

From \sax\ \citep[source 44,][]{2000ApJ...531L.131I} which 
is located at the rim of the field of \snrb, we detect pulsations with a period 
of 341.2 s. This indicates that the period of large spin-up reported by 
\citep{2000ApJ...531L.131I} is now lasting more than 4.5 years 
(Figs.~\ref{fig-spin} and \ref{fig-hist}). In Fig.~\ref{fig-hist} we also
plot period measurements of the two long period pulsars 
RX\,J0050.8$-$7316 and RX\,J0049.7$-$7323. The two pulsars with periods 
measured at more than two epochs, 
\sax\ and RX\,J0050.8$-$7316=AX\,J0051$-$733, show very similar pulse periods and 
also (linear) period changes $\dot{P}$ of $-$1.6 s yr$^{-1}$ and $-$1.0 s yr$^{-1}$, respectively.
A $\dot{P}$ of -14 s yr$^{-1}$ is derived for the 750 s pulsar 
XMMU\,004942.3$-$732313 but is based on only two period measurements, 
one of them with large error.

For six of the eleven sources we do not detect significant pulsations. From a 
Z$_1^2$ analysis no period with more than 2.5 $\sigma$ significance was found. In 
particular in the data of observation 0110000101 with low background, modulations 
with a similar pulsed fraction as seen from the detected pulsars (Fig.~\ref{fig-pow}) 
should be easily picked up. On the other hand, the sensitivity for period detection 
is lower in the high background observation 0110000301. The pulse period
of \sax\ was detected with a formal 3.2 $\sigma$ significance (for an unknown 
period) and a slightly weaker modulation would not be detected for sources 43 and 48. 

\subsection{X-ray spectra}

\begin{figure}
\begin{center}
\resizebox{8.6cm}{!}{\includegraphics[clip=,angle=-90]{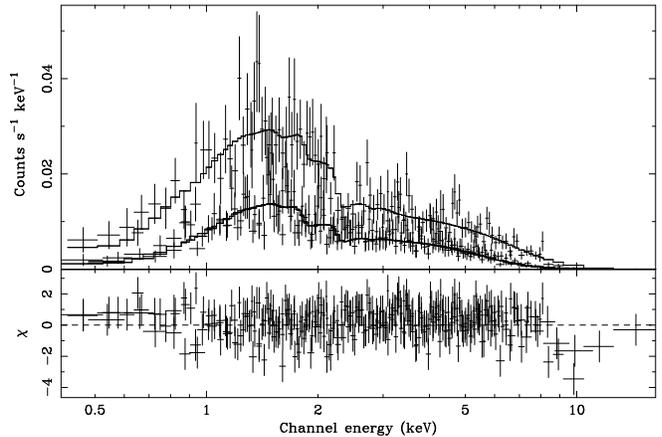}}
\end{center}
\caption{EPIC spectra (crosses) of the pulsar RX\,J0049.7-7323=AX\,J0049.5-7323
together with the best-fit absorbed power-law model. The model is drawn in the 
upper panel as histograms for the pn spectrum (upper one) and both MOS spectra
(lower pair which overlaps and appears as single histogram).}
\label{fig-spec}
\end{figure}

EPIC-pn and -MOS spectra were extracted for the eleven Be/X-ray binaries and candidates and
simultaneously fit. An absorbed power-law model describes the data sufficiently
well (with typical reduced $\chi^2$ values around 1.0) for the given statistical 
quality of the spectra. Normalization factors were allowed to vary between the 
individual instruments to account for cross-calibration effects. 
Fig.~\ref{fig-spec} shows an example fit to the spectra of XMMU\,J004942.3$-$732313
(source 11), the 750~s pulsar located in the field of \snra. 
Table~2 %Table~\ref{tab-fits} 
summarizes the spectral fit results for all eleven sources. 
The power-law photon indices $\gamma$ show a narrow distribution with an average 
value of 1.0 and a standard deviation of 0.16. The total absorption column densities 
are above the Galactic foreground value of 5.74\hcm{20} as observed towards 
the SMC \citep{1990ARAA...28..215D} and indicates absorption by interstellar 
matter in the SMC and/or local matter around the X-ray sources. For the 
absorbing material solar element abundances were used in the spectral fits.
If an SMC abundance of 0.2 solar is assumed \citep{1992ApJ...384..508R}, the 
contribution to the column density by SMC matter (total \nh\ reduced by the Galactic
foreground) needs to be multiplied by a factor of 5.0. In Table~2 %Table~\ref{tab-fits} 
we list observed fluxes and source intrinsic luminosities in the 0.2-10.0 keV 
energy band.

\subsection{Long term variability}

To investigate intensity variations on time scales of years we list in 
Table~2 %Table~\ref{tab-fits}
the range of X-ray luminosities in which the investigated 
sources were detected in the past. We do this separately for instruments covering 
an energy band similar to the EPIC band (ASCA, BeppoSAX) and for ROSAT with the soft band 
of 0.1-2.4 keV. Most broad band luminosities were reported from ASCA observations 
by \citet[][ 0.7-10.0 keV]{2003PASJ...55..161Y} which were corrected 
by a factor of 1.05 to account for the slightly larger (0.2-10.0 keV) energy band. Only 
source 44 (\sax) was observed with maximum luminosity by BeppoSAX 
\citep{2000ApJ...531L.131I}. In this case the correction factor is 1.23. The 
correction factors were derived for an unabsorbed power-law spectrum with 
photon index of 1.0. Source 5 was not detected by ASCA and for sources 7 and 9
the association with the nearby ASCA source 8 is unclear. We also do not give 
an ASCA flux for source 48 because of contamination of the spectrum by nearby 
sources due to the large ASCA point spread function. 
To obtain 0.2-10 keV luminosities based on ROSAT measurements we used the 
spectral parameters inferred from the EPIC spectra, computed expected PSPC count 
rates by folding the spectral model through the PSPC detector response and scaled the 
observed count rates to derive the conversion factor. Measured PSPC count rates were 
taken from the Results Archive (RRA) PSPC source catalogue from pointed ROSAT observations 
\citep{2000rra} which contains multiple detection entries for a source if observed more 
than once during the ROSAT mission. The last column in Table~2 %Table~\ref{tab-fits} 
shows the ratio between maximum and minimum 
luminosity which is at most a factor of 13. The intensity variations are relatively 
moderate for this kind of objects and can probably be explained by variations of 
the mass accretion rate along a neutron star orbit with moderate eccentricity 
within the circum-stellar matter of the Be star. While variations up to a factor 
of $\sim$20 are observed from many Be/X-ray binaries 
\citep[e.g.][]{1999A&A...344..521H}, `giant' outbursts (factor 100-1000) are rare 
and are thought to be explained by increased mass outflow through the Be stellar 
wind disc \citep{1994SSRv...69..255A}. No giant outbursts were so far detected from
any of the eleven systems.

% how often do Be stars show such outbursts

\section{The HMXB population of the SMC}

\begin{table*}
\begin{center}
\caption[]{High mass X-ray binaries and candidates in the SMC.}
\begin{tabular}{lrrrcrrrrlp{58mm}l}
\hline\noalign{\smallskip}
\hline\noalign{\smallskip}
\multicolumn{1}{l}{1} &
\multicolumn{1}{c}{2} &
\multicolumn{1}{c}{3} &
\multicolumn{1}{c}{4} &
\multicolumn{1}{c}{5} &
\multicolumn{1}{c}{6} &
\multicolumn{1}{c}{7} &
\multicolumn{1}{c}{8} &
\multicolumn{1}{c}{9} &
\multicolumn{1}{l}{10} &
\multicolumn{1}{l}{11} \\

\noalign{\smallskip}\hline\noalign{\smallskip}
\multicolumn{1}{l}{No} &
\multicolumn{1}{c}{RA} &
\multicolumn{1}{c}{Dec} &
\multicolumn{1}{c}{r$_{90}$} &
\multicolumn{1}{c}{S} &
\multicolumn{1}{c}{HS} &
\multicolumn{1}{c}{MA} &
\multicolumn{1}{c}{d$_{\rm ox}$} &
\multicolumn{1}{c}{Period} &
\multicolumn{1}{l}{ST} &
\multicolumn{1}{l}{Comment} \\

\multicolumn{1}{l}{} &
\multicolumn{2}{c}{(J2000.0)} &
\multicolumn{1}{c}{[\arcsec]} &
\multicolumn{1}{c}{} &
\multicolumn{1}{c}{} &
\multicolumn{1}{c}{} &
\multicolumn{1}{c}{[\arcsec]} &
\multicolumn{1}{c}{[s]} &
\multicolumn{1}{c}{} &
\multicolumn{1}{l}{} \\

\noalign{\smallskip}\hline\noalign{\smallskip}
~1     & 00 32 56.2 & $-$73 48 20 & 14.7 & R &  1 &	 &	 &	   & Be (2)      & RX\,J0032.9$-$7348 (1) two Be stars \\
~2     & 00 45 37.9 & $-$73 13 54 & 30.2 & R &  3 &  114 &  23.4 &	   & Be?         & RX\,J0045.6$-$7313 \\
~3$^a$ & 00 47 23.7 & $-$73 12 27 &  4.0 & N &  6 &  172 &   3.9 &  263	   & Be?         & RX\,J0047.3$-$7312 AX\,J0047.3$-$7312 (3) \\
~4$^b$ & 00 48 14.0 & $-$73 09 39 & 40.0 & A &	&  215 &  23.3 &	   & Be?         & AX\,J0048.2$-$7309 (3) \\
~5$^a$ & 00 48 34.5 & $-$73 02 30 &  4.1 & N &  8 &  238 &   2.2 &	   & Be?         & RX\,J0048.5$-$7302 \\
~6     & 00 49 02.6 & $-$72 50 53 & 15.4 & R &  9 &	 &	 &   74.67 & Be (2,5)    & RX\,J0049.1$-$7250 (1,4) AX\,J0049$-$729 (3) two Be stars \\
~7$^a$ & 00 49 13.8 & $-$73 11 37 &  4.0 & N &	&	 &	 &	   & Be?         & RX\,J0049.2$-$7311 (6) \\
~8$^b$ & 00 49 18.5 & $-$73 12 01 & 40.0 & A &	&	 &	 &    9.13 & Be?         & AX\,J0049$-$732 (3) \\
~9$^a$ & 00 49 29.9 & $-$73 10 58 &  4.1 & N & 10 &  300 &   1.7 &	   & Be?         & RX\,J0049.5$-$7310 (6) \\
10     & 00 49 33.7 & $-$73 31 25 & 24.2 & R & 11 &  302 &  22.2 &	   & Be?         & RX\,J0049.5$-$7331 AX\,J0049.5$-$7330 (3) \\
11$^a$ & 00 49 42.4 & $-$73 23 13 &  4.0 & N & 12 &  315 &   2.6 &  755.5  & Be (7)      & RX\,J0049.7$-$7323=AX\,J0049.5$-$7323 (3) \\
12$^a$ & 00 50 45.2 & $-$73 16 03 &  4.0 & N & 13 &  387 &   3.0 &  323.2  & Be (5,8,37) & RX\,J0050.8$-$7316=AX\,J0051$-$733 (8,3) \\
13     & 00 50 46.9 & $-$73 32 48 & 34.4 & R & 14 &  393 &   9.2 &	   & Be?         & RX\,J0050.7$-$7332 \\
14$^a$ & 00 50 57.6 & $-$73 10 08 &  4.0 & N & 15 &  414 &   1.7 &	   & Be?         & RX\,J0050.9$-$7310 AX\,J0050.8$-$7310 (3) \\
15     & 00 50 56.9 & $-$72 13 31 &  7.1 & R & 16 &  413 &   2.7 &   91.12 & Be (2)      & AX\,J0051$-$722 (9) \\
16     & 00 51 19.6 & $-$72 50 44 & 17.1 & R & 17 &  447 &  14.6 &	   & Be?         & RX\,J0051.3$-$7250 \\
17$^a$ & 00 51 52.3 & $-$73 10 33 &  4.0 & N & 19 &  504 &   2.1 &  172	   & Be (8)      & RX\,J0051.9$-$7311=AX\,J0051.6$-$7311 (8,3) \\
18     & 00 51 53.1 & $-$72 31 50 & 46.3 & R & 20 &  506 &   1.1 &    8.9  & Be (2,12)   & 1WGA\,J0051.8$-$7231 (1,12) \\
19     & 00 51 54.2 & $-$72 55 36 & 40.0 & E & 21 &  521 &  28.3 &	   & Be?         & (13) \\
20     & 00 52 05.4 & $-$72 25 56 & 17.3 & R & 22 &  531 &   7.2 &	   & Be          & SMC X$-$3 \\
21     & 00 52 14.0 & $-$73 19 14 &  7.3 & R & 23 &  552 &   5.1 &   15.3  & Be (14,15)  & RX\,J0052.1$-$7319 (1,14) \\
22     & 00 52 17.0 & $-$72 19 51 &111.0 & X &	  &  537 &  71.1 &    4.78 & Be? (19)    & XTE\,J0052$-$723 (19) \\
23     & 00 52 52.7 & $-$72 48 22 & 40.0 & E & 24 &  618 &   8.1 &	   & Be (20)     & 2E0051.1$-$7304, AzV138 \\
24     & 00 52 57.9 & $-$71 58 13 & 31.0 & R & 25 &  623 &  13.2 &  169	   & Be (8)      & RX\,J0052.9$-$7158 (8) XTE\,J0054$-$720 =AX\,J0052.9$-$7157 (10,3) 					   \\
25     & 00 53 30.0 & $-$72 27 28 & 27.8 & R & 26 &  667 &  31.0 &	   & Be?         & \\
26     & 00 53 53.4 & $-$72 27 01 & 26.6 & R & 27 &  717 &  22.7 &   46.63 & Be (11)     & 1WGA\,J0053.8$-$7226=XTE\,J0053$-$724 =AX\,J0053.9$-$7226 (9,3) two Be stars \\
27     & 00 54 30.9 & $-$73 40 55 & 12.6 & R & 28 &      &       &    2.37 & Be          & SMC X$-$2 (38,3) \\
28     & 00 54 33.2 & $-$72 28 09 & 46.2 & R & 29 &  772 &  26.6 &	   & Be?         & \\
29     & 00 54 55.4 & $-$72 45 06 &  9.0 & R & 30 &  809 &   4.7 &	   & Be?         & AX\,J0054.8$-$7244 (3) \\
30     & 00 54 56.0 & $-$72 26 49 &  4.1 & N & 31 &  810 &   1.6 &   59.07 & Be (2,5)    & XTE\,J0055$-$724=1SAX\,J0054.9$-$7226 =1WGA\,J0054.9$-$7226 (21,22,23) \\
31     & 00 56 05.2 & $-$72 22 01 &  4.5 & N & 32 &  904 &   3.2 &  140.1  & Be?         & XMMU\,J005605.2$-$722200 2E0054.4$-$7237 (23) \\
32     & 00 57 26.8 & $-$73 25 02 & 21.1 & R &	&      &       &  101	   & Be?         & RX\,J0057.4$-$7325 AX\,J0057.4$-$7325 (3) (7) \\
33     & 00 57 36.3 & $-$72 19 35 &  1.1 & C &	& 1020 &   1.6 &  565	   & Be?         & CXOU\,J005736.2$-$721934 (23,24)  \\
34     & 00 57 50.2 & $-$72 02 37 &  4.1 & N & 35 & 1036 &   3.5 &  280.4  & Be?         & RX\,J0057.8$-$7202=AX\,J0058$-$72.0 (3,23) \\
35     & 00 57 50.3 & $-$72 07 57 &  1.0 & C & 36 & 1038 &   0.7 &  152.3  & Be?         & CXOU\,J005750.3$-$720756 (23,24) \\
36     & 00 57 59.5 & $-$71 56 37 & 20.4 & R & 37 & 1044 &  21.1 &	   & Be?         & \\
37     & 00 58 11.7 & $-$72 30 50 &  4.3 & N & 38 &      &       &	   & Be (7,8)    & RX\,J0058.2$-$7231 (8,23) \\
38     & 00 59 11.4 & $-$71 38 45 &  7.5 & R & 40 & -179 & 10.1  &   2.763 & Be (26)     & RX\,J0059.2$-$7138 (25) \\
39     & 01 00 30.2 & $-$72 20 35 &  5.1 & N &	  & 1208 &   2.1 &         & Be?         & XMMU\,J010030.2$-$722035 (23) \\
40     & 01 01 02.8 & $-$72 06 58 &  1.2 & C & 42 & 1240 &   1.1 &  304.5  & Be (2,7)    & RX\,J0101.0$-$7206 (1,24) \\
41     & 01 01 20.8 & $-$72 11 21 &  4.1 & N & 43 & 1257 &   2.1 &  455	   & Be (27)     & RX\,J0101.3$-$7211 (27) \\
42     & 01 01 37.6 & $-$72 04 19 &  4.1 & N & 44 & 1277 &   5.2 &	   & Be?         & RX\,J0101.6$-$7204 (23)\\
43$^a$ & 01 01 52.4 & $-$72 23 36 &  4.1 & N & 45 & 1288 &   2.8 &	   & Be?         & AX\,J0101.8$-$7223\,(3) XMMU\,J010152.4$-$722336 \\
44$^a$ & 01 03 14.2 & $-$72 09 15 &  4.1 & N & 49 & 1367 &   1.1 &  345.2  & Be (29,5)   & SAX\,J0103.2$-$7209=AX\,J0103.2$-$7209 =CXOU\,J010314.1$-$720915 (28) \\
45     & 01 03 37.6 & $-$72 01 33 &  4.0 & N & 50 & 1393 &   1.1 &	   & Be?         & RX\,J0103.6$-$7201 (23) \\
46     & 01 04 07.4 & $-$72 43 59 & 19.0 & R & 51 & 1440 &   9.0 &	   & Be?         & or AGN? 13cm  \\
47$^b$ & 01 04 35.7 & $-$72 21 43 &  7.4 & R & 52 & 1470 &   4.0 &	   & Be?         & RX\,J0104.5$-$7221 \\
48$^a$ & 01 05 09.8 & $-$72 11 46 &  4.2 & N & 53 &      &       &   3.34? & Be?         & RX\,J0105.1$-$7211 AX\,J0105$-$722 (3,30) \\
\noalign{\smallskip}\hline\noalign{\smallskip}
\end{tabular}
\end{center}
\label{tab-smc}
\end{table*}

\addtocounter{table}{-1}
\begin{table*}
\label{tab-cont}
\begin{center}
\caption[]{Continued.}
\begin{tabular}{lrrrcrrrrlp{58mm}l}
\hline\noalign{\smallskip}
\hline\noalign{\smallskip}
\multicolumn{1}{l}{No} &
\multicolumn{1}{c}{RA} &
\multicolumn{1}{c}{Dec} &
\multicolumn{1}{c}{r$_{90}$} &
\multicolumn{1}{c}{S} &
\multicolumn{1}{c}{HS} &
\multicolumn{1}{c}{MA} &
\multicolumn{1}{c}{d$_{\rm ox}$} &
\multicolumn{1}{c}{Period} &
\multicolumn{1}{l}{ST} &
\multicolumn{1}{l}{Comment} \\

\multicolumn{1}{l}{} &
\multicolumn{2}{c}{(J2000.0)} &
\multicolumn{1}{c}{[\arcsec]} &
\multicolumn{1}{c}{} &
\multicolumn{1}{c}{} &
\multicolumn{1}{c}{} &
\multicolumn{1}{c}{[\arcsec]} &
\multicolumn{1}{c}{[s]} &
\multicolumn{1}{c}{} &
\multicolumn{1}{l}{} \\

\noalign{\smallskip}\hline\noalign{\smallskip}
49     & 01 05 55.4 & $-$72 03 48 &  4.3 & N & 55 & 1557 &   3.2 &	   & Be?         & RX\,J0105.9$-$7203  AX\,J0105.8$-$7203 (3,23) \\
50     & 01 07 10.9 & $-$72 35 37 & 13.1 & R & 56 & 1619 &  10.1 &	   & Be?         & AX\,J0107.2$-$7234 (3) \\
51     & 01 11 14.5 & $-$73 16 50 & 61.2 & R & 58 & 1747 &  63.9 &   31.03 & Be (15,16)  & XTE\,J0111.2$-$7317=AX\,J0111.1$-$7316 (17,18,3) \\
52     & 01 17 05.5 & $-$73 26 33 &  7.0 & R &    &	 &	 &  0.717  & SG          & SMC X$-$1 \\
53     & 01 17 41.5 & $-$73 30 50 &  7.0 & R & 59 & 1845 &   4.3 &   22.07 & Be (32)     & RX\,J0117.6$-$7330 (31) \\
54     & 01 19 37.7 & $-$73 30 06 & 14.7 & R & 60 & 1867 &   8.3 &	   & Be?         & \\
\\
55     &	    &	          &	 & X &    &	 &	 &   2.165 &		 & XTE\,J0119-731 (40) \\
56     &	    &	          &	 & X &    &	 &	 &   6.85  &		 & XTE\,J0103-728 (39) \\
57     &	    &	          &	 & X &    &	 &	 &   7.8   &		 & (36) \\
58     &	    &	          &	 & X &    &	 &	 &   16.6  &		 & XTE\,J0050$-$732\#1 (35) $\neq$RX\,J0051.9$-$7311 \\
59     &	    &	          &	 & X &    &	 &	 &   25.5  &		 & XTE\,J0050$-$732\#2 (35)\\
60     &	    &	          &	 & X &    &	 &       &   46.4  &             & XTE pulsar (33) \\
61     &	    &	          &	 & X &    &	 &	 &   82.4  &             & XTE\,J0052$-$725 (33) \\
62     &	    &	          &	 & X &    &	 &	 &   89    &		 & (39) \\
63     &	    &	          &	 & X &    &	 &	 &   95    &             & XTE SMC95 (34) \\
64     &	    &	          &	 & X &    &	 &	 &   144.1 &		 & XTE SMC144s (39) \\
65     &	    &	          &	 & X &    &	 &	 &   164.7 &		 & (39) \\
\noalign{\smallskip}\hline\noalign{\smallskip}
\end{tabular}
\end{center}
$^a$ Source investigated in this work. ~~~ $^b$ Source not detected by XMM-Newton in this work. \\
For sources 55-65 no X-ray positions are listed due to their large uncertainties and we list them sorted by pulse period.\\
Notes to specific columns: (2,3,4) Best available X-ray position with error.
For ROSAT and XMM-Newton positions systematic uncertainties of 7\arcsec\ and 4\arcsec\ are included.
(5) X-ray mission which provided the data for the best
position (A: ASCA, C: Chandra, E: Einstein, N: XMM-Newton, R: ROSAT, X: RXTE).
(6) as Table~1. (7) Nearest \Halp\ emission line object consistent with the X-ray position from the catalogues 
of \citet{1993A&AS..102..451M} and \citet[][ in the case of source 38 which is indicated 
by a negative entry number]{2000MNRAS.311..741M}. (8) as Table~1.
(10) HMXB sub-type, Be or supergiant (SG); Be? indicates an \Halp\ emission line 
object as possible counterpart, no optical spectrum is available in the literature in 
this case.
(11) Likely cross identifications of X-ray sources 
detected with different missions (a `=' between source names indicates a secure 
identification based on positional coincidence and detection of pulsations, the 
$\neq$ in the comment to source 58 marks that the ROSAT source is not the counterpart of
the RXTE source, see also source 17). \\
References given in parenthesis in columns 10 and 11:\\
(1) \citet{1996A&A...312..919K}; 
(2) \citet{1999MNRAS.309..421S};
(3) \citet{2003PASJ...55..161Y};
(4) \citet{1998IAUC.6840....1K};
(5) \citet{2000MNRAS.311..169C};
(6) \citet{2000A&A...361..823F};
(7) \citet{2003MNRAS.338..428E};
(8) \citet{1999AJ....117..927S};
(9) \citet{1998IAUC.6803....1C};
(10) \citet{1998IAUC.6814....1L};
(11) \citet{2001MNRAS.320..281B};
(12) \citet{1997ApJ...484L.141I};
(13) \citet{1992ApJS...78..391W};
(14) \citet{2001ApJ...560..378F};
(15) \citet{2001A&A...374.1009C};
(16) \citet{2000MNRAS.314..290C};
(17) \citet{1998IAUC.7048....1C};
(18) \citet{1998IAUC.7048....2W};
(19) \citet{2003MNRAS.339..435L};
(20) \citet{1985AJ.....90.2009G};
(21) \citet{1998IAUC.6818R...1M};
(22) \citet{1998IAUC.6818R...2S};
(23) \citet{2003A&A...403..901S};
(24) \citet{2003ApJ...584L..79M};
(25) \citet{1994ApJ...427L..25H};
(26) \citet{1996MNRAS.281L..63S};
(27) \citet{2001A&A...369L..29S};
(28) \citet{2000ApJ...531L.131I};
(29) \citet{1994AJ....107.1363H};
(30) \citet{2000A&A...353..129F};
(31) \citet{1999ApJ...518L..99M};
(32) \citet{1997ApJ...474L.111C};
(33) \citet{2002IAUC.7932....2C};
(34) \citet{2002A&A...385..464L};
(35) \citet{2002ApJ...567L.129L};
(36) \citet{2003HEAD...35.1730C};
(37) \citet{2002MNRAS.332..473C};
(38) \citet{2001ApJ...548L..41C};
(39) \citet{2003ATel..163....1C};
(40) \citet{2003IAUC.8064....4C};
\end{table*}

The detection of a periodic 263.6~s X-ray flux modulation from 
XMMU\,J004723.7$-$731226 increases the number of likely HMXB pulsars in the SMC 
to at least thirty-seven. In Table~\ref{tab-smc} we give an updated list of HMXBs 
and candidates in the SMC which is based on the work of HS00
and add transient pulsars detected by RXTE. We do not include two additional known pulsars which may be of different type:
AX\,J0043$-$737 for which confirmation of a 87 ms period is needed and which could be a Crab-like 
pulsar \citep{2000IAUC.7361....2Y,2003PASJ...55..161Y}, and CXOU\,J0110043.1$-$721134
a possible anomalous X-ray pulsar in the SMC \citep{2002ApJ...574L..29L}.

\begin{figure}
\begin{center}
\unitlength=1cm
\begin{picture}(8.6,6.0)
\put(0.0,6.0){\resizebox{8.6cm}{!}{\includegraphics[clip=,angle=-90]{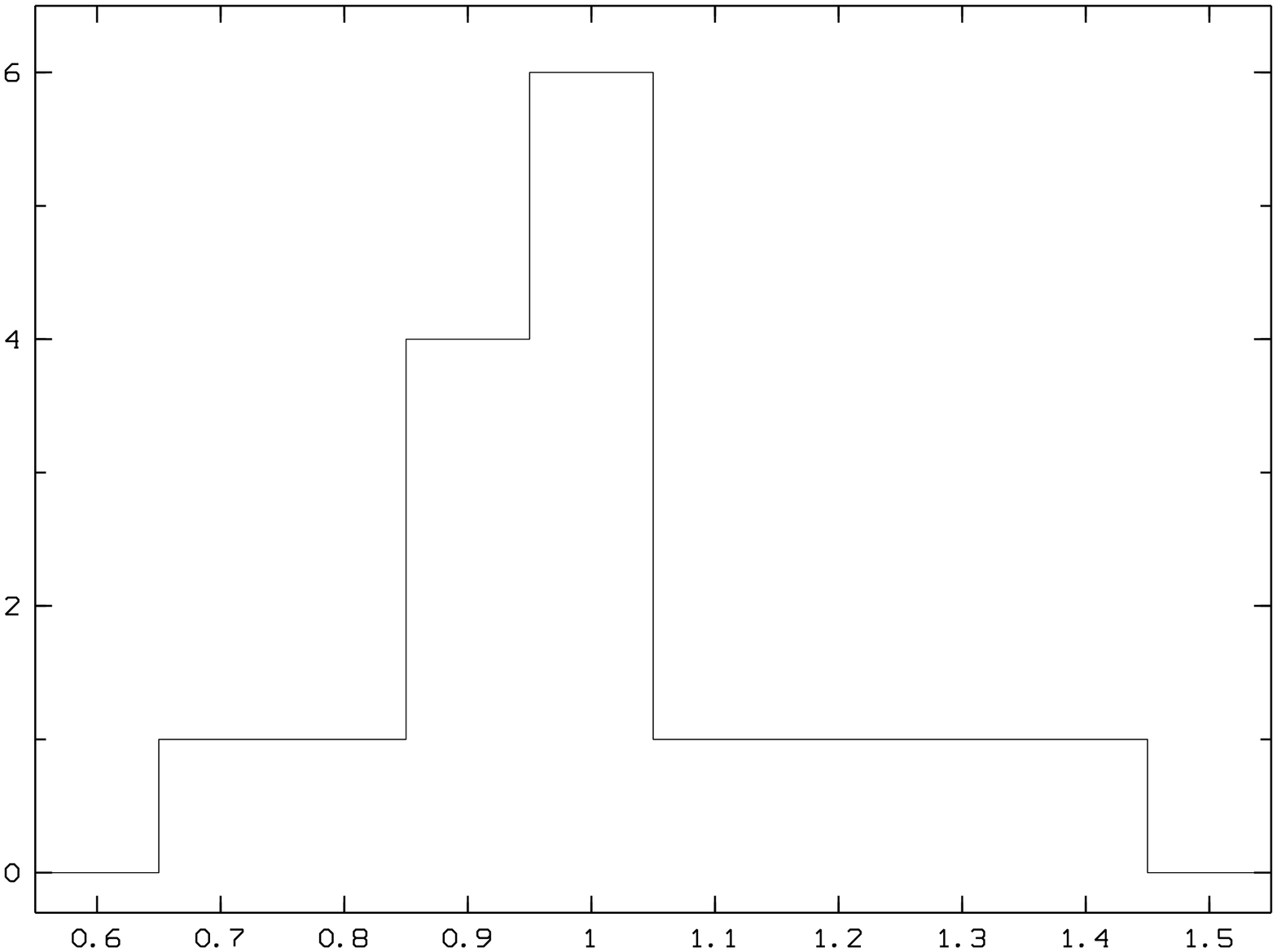}}}
\put(0.5,2.3){\scalebox{1.0}{\rotatebox{90}{Frequency}}}
\put(3.7,0.0){\scalebox{1.0}{\rotatebox{00}{Photon index}}}
\end{picture}
\begin{picture}(8.6,6.0)
\put(0.0,6.0){\resizebox{8.6cm}{!}{\includegraphics[clip=,angle=-90]{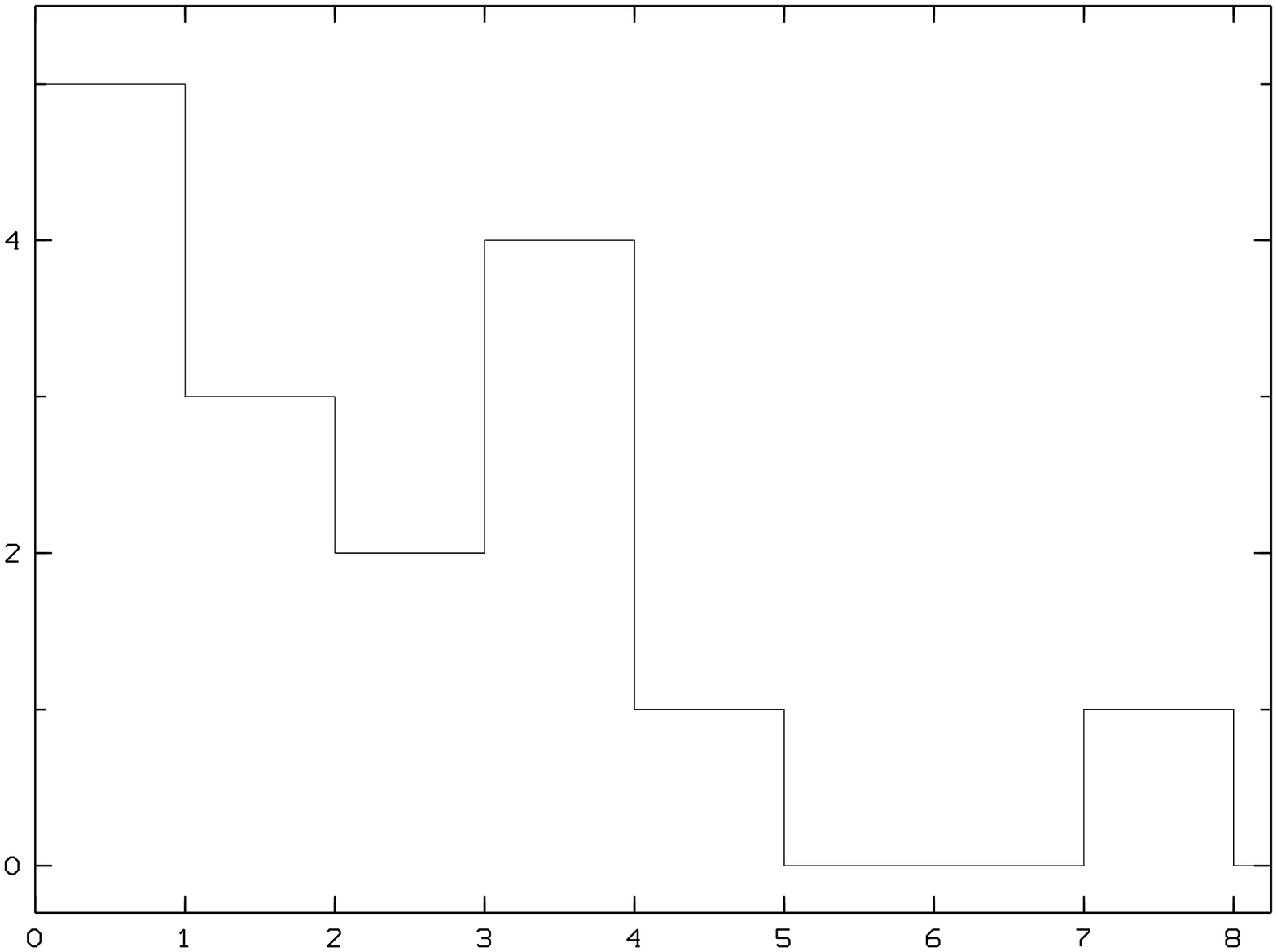}}}
\put(0.5,2.3){\scalebox{1.0}{\rotatebox{90}{Frequency}}}
\put(2.8,0.0){\scalebox{1.0}{\rotatebox{00}{Column density (\ohcm{21})}}}
\end{picture}
\end{center}
\caption{Distribution of power-law photon indices (top) and absorption column density 
(bottom, without Galactic absorption) measured from EPIC spectra of sixteen Be/X-ray binaries
and candidates in the SMC observed by the EPIC instruments of XMM-Newton.}
\label{fig-histo}
\end{figure}

The number of Be/X-ray binaries and candidates is the  highest known from any galaxy, 
including the Milky Way where about forty such systems \citep{2000A&AS..147...25L} are
known today. The SMC Be/X-ray binaries, all at a similar distance, constitute an ideal 
sample for statistical studies. Luminosity distributions were derived by HS00 and 
\citet{2003A&A...403..901S}. Compared to the Milky Way the SMC harbours a higher fraction 
of low-luminosity systems ($\sim$\oergs{35}), which may be a selection effect due to
a more difficult identification of such objects in the Galactic Plane (HS00).
The XMM-Newton results from SMC HMXBs allow to compare the X-ray spectra from already 
16 sources [11 from this work and sources 33, 34, 35, 41 and 45 from \citet{2003A&A...403..901S}] 
obtained with the same instruments and 
processed in a coherent way. In Fig.~\ref{fig-histo} (upper panel) the distribution of the 
power-law photon index is presented. The distribution is strongly peaked at 1.0, but includes 
a weak wide contribution with indices between 0.65 and 1.45. 

Although the column densities derived from the spectra have relatively large errors, 
the \nh\ distribution (Fig.~\ref{fig-histo} bottom panel) exhibits two maxima. 
An excess of objects with little or no absorption may either indicate spectra with a soft 
component which is not recognized because of insufficient statistics
or a special spatial distribution of the sources on the near side of the SMC (star formation arm). 
The second peak may be explained by interstellar absorption
in the SMC with the majority of objects located in the inner parts of the SMC. A comparison with
similar data obtained from background AGN behind the SMC should allow to derive constraints on 
the total absorbing column density through the SMC on a statistical basis.

\section{Summary}

We performed temporal and spectral analyses of eleven Be/X-ray binaries and candidates observed
serendipitously by the EPIC instruments on board XMM-Newton in two observations of the supernova 
remnants \snra\ and \snrb. From five of the objects pulsations in their X-ray flux are detected,
four of them were previously known pulsars and in one case, XMMU\,J004723.7$-$731227, the XMM-Newton 
data revealed 263 s pulsations for the first time. XMMU\,J004723.7$-$731227 coincides in position with 
RX\,J0047.3$-$7312 which was suggested to be a Be/X-ray binary due to its likely identification with
an \Halp\ emission line object (HS00). The detection of the pulse period strongly supports this 
identification. 

The detection of pulsations, together with the good X-ray positions allowed to securely identify
sources detected by various X-ray space missions, like e.g. the 172 s pulsar AX\,J0051.6$-$7311 with 
XMMU\,J005152.2$-$731033 (via pulsations) and RX\,J0051.9$-$7311 (via accurate position). The latter
was previously also suggested to be the counterpart of XTE\,J0050$-$732\#1, a 16.6 s pulsar.
The 345 s pulsar \sax\ was found to continue its large spin down with a rate of $-$1.6 s yr$^{-1}$ 
extending now over 4.5 years. The 323 s pulsar AX\,J0051$-$733 was found to exhibit very similar 
properties as \sax\ with respect to spin period, period derivative and X-ray luminosity.

The large number of Be/X-ray binaries in the SMC allows first statistical studies of their
spectral and temporal properties. XMM-Newton and Chandra are ideally suited to detect pulsars
among the Be-/X-ray binary candidates at low source fluxes and future observations are required 
to further increase the sample and allow to analyse it in a homogeneous way.

\begin{acknowledgements}
The XMM-Newton project is supported by the Bundesministerium f\"ur Bildung und
For\-schung / Deutsches Zentrum f\"ur Luft- und Raumfahrt (BMBF / DLR), the
Max-Planck-Gesellschaft and the Heidenhain-Stif\-tung. 
\end{acknowledgements}

\bibliographystyle{apj}
\bibliography{mcs,general,myrefereed,myunrefereed,mytechnical}

\end{document}